\documentclass{article}
\usepackage{epsfig}
\title{Quantum walks based on an interferometric analogy}
\author{Mark Hillery$\,^{1}$, Janos Bergou$\,^{1}$ and Edgar Feldman$\,^{2}$ 
\\ $\,^{1}$ Department of Physics, Hunter College of CUNY \\ 695 Park 
Avenue, New York, NY 10021 \\ \\ $\,^{2}$ Department of Mathematics \\ 
Graduate Center of the City University of New York \\ 365 Fifth Avenue, 
New York, NY 10016}
\begin{document}
\maketitle
\abstract{There are presently two models for quantum walks on graphs.  
The ``coined'' walk uses discrete time steps, and contains, besides 
the particle making the walk, a second quantum system, the coin, that
determines the direction in which the particle will move.  The
continuous walks operate with continuous time.  Here a third model
for quatum walks is proposed, which is based on an analogy to optical
interferometers.  It is a discrete-time model, and the unitary
operator that advances the walk one step depnds only on the local
structure of the graph on which the walk is taking place.  This type
of walk also allows us to introduce elements, such as phase shifters,
that have no counterpart in classical random walks.  Several
examples are discussed.}
\section{Introduction}
Random walks on graphs are the basis of a number of classical algorithms,
Examples include, 2-SAT, graph connectivity, and finding satisfying
assignments for Boolean formulas.  As a result, it is natural to 
explore the quantum counterpart of a random walk, in the hope that it
will be useful in the development of quantum algorithms.  This has
led to a number of studies.  Quantum walks on the line were
examined by Nayak and Vishwananth \cite{nayak}, and on the cycle by 
Aharonov, et al. \cite{aharonov}.  The latter study also considered
a number of properties of quantum walks on general graphs.  Numerical 
simulations of walks in two and three dimensions were performed by
Mackay, et al. \cite{mackay}.  Absorbing times and probabilities of
quantum walks on the line were studied by several authors 
\cite{yamasaki,bach}.  One of the main results to come from this work is
that quantum walks spread faster than do classical ones.  In particular,
on the line, the standard deviation of the position of the particle
making the walk increases linearly with the number of steps rather
that with its square root as in the classical case.  Walks on the 
hypercube have also been considered, and here the results are
even more dramatic \cite{moore,kempe}.  Kempe has shown that the
hitting time for the walk from one corner of an $n$-bit hypercube
to the opposite corner is polynomial in $n$ for a quantum walk,
but exponential for a classical one. The quantum walk on the 
hypercube was subsequently used as the basis of a quantum search
algorithm \cite{kempe2}.  The effect of decoherence on
these walks has also been studied. Brun et al.\ showed how increasing
dedoherence turns a quantum walk into a classical random walk 
\cite{brun}.   Kendon and Treganna found that small amounts of 
decoherence can actually speed the convergence of the time-averaged
probability distribution of a particle in a quantum walk to a
uniform distribution \cite {kendon}. 

The time steps in the quantum walks considered in these works are discrete.  
Continuous-time quantum walks have also been proposed \cite{childs}.  It
was shown that on a particular graph, the propagation between two
properly chosen nodes is exponentially faster in the quantum case.

There is, at the moment, only one algorithm based on quantum walks on
graphs, and it was proposed very recently by Childs, et al.\ \cite{childs2}.
They constructed an oracle problem that can be solved by exponentially
faster on a quantum computer than on a classical one by utilizing 
continuous quantum walks.  The vertices of the graph are numbered (named),
and there are two special vertices called the entrance and the exit.  The 
problem is, given the name of the entrance, and the oracle, find the name
of the exit.  The oracle specifies the graph, which belongs to a particular 
set of possible graphs, by taking a binary number as its input, and either
telling you that this number does not correspond to a vertex, or if it 
does, telling you the names of the adjacent vertices. 

All of the discrete-time quantum walks are based on a particular model,
the, ``coined quantum walk'', due to Watrous \cite{watrous}.  
In trying to formulate
a quantum walk on a graph, the most natural thing to do is to let a
set of orthonormal basis states correspond to the vertices of the graph.
If a particle is in the state $|n\rangle$, that corresponds to its being
located on vertex $n$.  Trying to define a unitary evolution using this
scheme soon leads to serious problems, as was first noted by Meyer
\cite{meyer}.  Watrous solved this problem by enlarging the Hilbert
space in which the quantum walk takes place.  How this scheme works
is most easily seen by considering the quantum walk on a line.  The
vertices are labelled by integers, and, in addition, there is a quantum
coin, which has two states, $|L\rangle$ and $|R\rangle$, corresponding
to left and right, respectively.  A basis for the Hilbert space describing
this system is given by the states $|n\rangle\otimes\alpha\rangle$, where 
$n$ is an integer, and $\alpha$ is either $L$ or $R$.  A step in this walk 
consists of applying the Hadamard operator, $H$, to the coin,
\begin{eqnarray}
H|L\rangle & = & \frac{1}{\sqrt{2}}(|L\rangle + |R\rangle ) \nonumber \\  
H|R\rangle & = & \frac{1}{\sqrt{2}}(|L\rangle - |R\rangle ) .
\end{eqnarray}
and then the operator 
\begin{equation}
V_{H} = S\otimes |R\rangle\langle R|+ S^{\dagger}\otimes |L\rangle\langle L|,
\end{equation}
where $S$ is the shift operator, whose action is given by
\begin{equation}
S|n\rangle = |n+1\rangle \hspace{1cm} S^{\dagger}|n\rangle = |n-1\rangle .
\end{equation}

The coined quantum walk can be extended in a simple way to regular
graphs, i.e.\ those in which all vertices have the same number of
edges emanating from them.  When this is not true, things become
more complicated, and it seems to be necessary to consdier the 
global structure of the graph in defining the walk.  So far, no studies
of discrete-time quantum walks on graphs that are not regular have
appeared.

What we wish to propose here is a different type of discrete-time
quantum walk.  It is based on thinking about the graph as an
interferometer.  The vertices are optical elements known as $2N$-ports,
where $N$ is the number of edges meeting at the vertex, and the edges
correspond to paths a photon can take through the interferometer.  There
is no quantum coin in these walks.  The states are labelled by the
edges rather than the vertices in the graph, and each edge has
two states.  If the edge is labelled $ab$, $a$ corresponding to one
end and $b$ to the other, then one state is $ab$, corresponding
to a photon going from $a$ to $b$, and the other is $ba$, corresponding
to a photon going from $b$ to $a$. This approach is easily extended
to arbitrary graphs; one simply writes down the transition rules for
each vertex, and all of them taken together define a unitary operator
that advances the walk one step.  In addition, we can add elements to
this walk that correspond to the addition of phase shifters to paths
in an interferometer.

This model of a quantum walk on
a graph is closely related to the optical networks considered by
T\"{o}rm\"{a} and Jex \cite{torma}. They considered two-dimensional
arrays of beam splitters and the propagation of photons through
them.  The horizontal motion of the photon in these networks
corresponds to the time steps in a quantum walk, and the vertical
position of the photon is just the position of the particle in the
quantum walk.  Note that these networks provide one with the opportunity
to simulate the model of quantum walks proposed here with linear
optics.

\section{Walk on the cycle}
Perhaps the simplest walk is that on a cycle or ring.  Let us label the
vertices by the numbers $0$ through $N-1$, where the vertex $N$ is
identical to $0$.  That is, if we move one step forward from the
vertex $N-1$, we end up at vertex $0$.  The states of the system are
$|j,k\rangle$, where $k=j\pm 1$, which can be thought of as a photon 
on the edge between vertices $j$ and $k$ going from $j$ to $k$.
Because each edge has two states, and there are $N$ edges, the dimension
of this space is $2N$.

The vertices can be thought of as beam splitters.  Consider what happens
when a photon travelling in the horizontal direction hits a vertical
beam splitter.  The photon has a certain amplitude to continue in
the direction it was going, i.e.\ to be transmitted, and an amplitude
to be change its direction, i.e.\ to be reflected.  The beam
splitter has two input modes, the photon can enter from either the
right or the left, and two output modes, the photon can leave heading
either right or left.  The beam splitter defines a unitary transformation
between the input and output modes.

We now need to translate this analogy into transition rules for our
quantum walk.  Suppose we are in the state $|j-1,j\rangle$.  If the
photon is transmitted it will be in the state $|j,j+1\rangle$, and if
reflected in the state $|j,j-1\rangle$.  Let the transmission amplitude
be $t$, and the reflection amplitude be $r$.  We then have the
transition rule
\begin{equation}
|j-1,j\rangle\rightarrow t|j,j+1\rangle +r|j,j-1\rangle ,
\end{equation}
where unitarity implies that $|t|^{2}+|r|^{2}=1$.
The other possibility is that the photon is incident on vertex
$j$ from the right, that is it is in the state $|j+1,j\rangle$.
If it is transmitted it is in the state $|j,j-1\rangle$, and if
it is reflected, it is in the state $|j,j+1\rangle$.  Unitarity
of the beam splitter transformation then gives us that
\begin{equation}
|j+1,j\rangle\rightarrow t^{\ast}|j,j-1\rangle -r^{\ast}|j,j+1\rangle .
\end{equation}
These rules specify our walk.

The case $t=1$ and $r=0$ corresponds to free particle propagation; a
``photon'' in the state $|j,j+1\rangle$ simply moves one step to the
right with each time step in the walk.  If $r\neq 0$, then there is
some amplitude to move both to the right and to the left.  A physical
system to which this is analogous is the motion of a particle in a
periodic potential.  The beam splitters can be thought of as scattering
centers with the scattering resulting from a localized potential.  As
is well known, this leads to energy bands, and, as we shall soon see,
a similar structure emerges in quantum walks on the cycle.

One way of approaching the study of the dynamics generated by this walk is
to find the eigenvalues and eigenstates of the unitary transformation, $U$, 
that moves the system a single step.  In order to do this, we first note
that $U$ commutes with the translation operator, $T$, where 
\begin{equation}
T|j,j+1\rangle = |j+1,j+2\rangle \hspace{1cm} T|j+1,j\rangle
=|j+2,j+1\rangle .
\end{equation}
This implies that these operators can be simultaneously diagonalized.
The eigenvalues of $T$ are $e^{i\theta_{k}}$, where $\theta_{k}=2\pi k/N$,
and $k=0,1,\ldots N-1$.  Each of these eigenvalues is doubly degenerate,
and the two dimensional space of eigenvectors corresponding to 
$e^{i\theta_{k}}$ is spanned by
\begin{eqnarray}
|u_{k+}\rangle & = & \frac{1}{\sqrt{N}}\sum_{j=0}^{N-1}e^{ij\theta_{k}}
|j,j+1\rangle \nonumber \\ 
|u_{k-}\rangle & = & \frac{1}{\sqrt{N}}\sum_{j=0}^{N-1}e^{ij\theta_{k}}
|j+1,j\rangle  .
\end{eqnarray}
The eigenstates of $U$ are just linear combinations of $|u_{k+}\rangle$
and $|u_{k-}\rangle$.  Defining
\begin{equation}
|\psi_{k}\rangle = a_{k+}|u_{k+}\rangle +a_{k-}|u_{k-}\rangle ,
\end{equation}
we find that the equation $U|\psi_{k}\rangle = \lambda |\psi_{k}\rangle$
becomes
\begin{equation}
\left( \begin{array}{cc} te^{-i\theta_{k}} & -r^{\ast} \\ r & t^{\ast}
e^{i\theta_{k}} \end{array} \right) \left( \begin{array} {c} a_{k+} \\
a_{k-} \end{array}\right) = \lambda \left( \begin{array} {c} a_{k+} \\
a_{k-} \end{array}\right) .
\end{equation}
Expressing $t$ as $t=|t|e^{i\eta}$, we find that the eigenvalues are
\begin{equation}
\label{eigenvalue}
\lambda_{k\pm}=|t|\cos (\theta_{k}-\eta )\pm i[1-|t|^{2}\cos^{2}(
\theta_{k}-\eta )]^{1/2} ,
\end{equation}
and the corresponding eigenfunctions are given by 
\begin{eqnarray}
a_{k+}^{(+)}& = & \frac{r^{\ast}}{[2C_{k}(C_{k}+S_{k})]^{1/2}} \nonumber \\
a_{k-}^{(+)}& = & \frac{-i(S_{k}+C_{k})}{[2C_{k}(C_{k}+S_{k})]^{1/2}} ,
\end{eqnarray}
for $\lambda_{k+}$, and 
\begin{eqnarray}
a_{k+}^{(-)}& = & \frac{r^{\ast}}{[2C_{k}(C_{k}-S_{k})]^{1/2}} \nonumber \\
a_{k-}^{(-)}& = & \frac{i(C_{k}-S_{k})}{[2C_{k}(C_{k}-S_{k})]^{1/2}} ,
\end{eqnarray}
for $\lambda_{k-}$.
Here we have defined
\begin{eqnarray}
C_{k} & = & [1-|t|^{2}\cos^{2}(\theta_{k}-\eta )]^{1/2} \nonumber \\
S_{k} & = & |t|\sin (\theta_{k} -\eta ) .
\end{eqnarray}

One thing we notice immediately, is that for all of these eigenstates,
the probability to be located on an edge, is the same for all edges,
just $1/N$.  That means, that for any initial state, $|\Psi_{in}\rangle$,
the average probability distribution, 
\begin{equation}
p_{j}^{(m)}=\frac{1}{m}\sum_{k=0}^{m-1}(|\langle j,j+1|U^{k}|\Psi_{in}
\rangle |^{2}+|\langle j+1,j|U^{k}|\Psi_{in}\rangle |^{2} ,
\end{equation}
where $p_{j}^{(m)}$ is the average probability of being on the edge between
$j$ and $j+1$ after $m$ steps, goes to a constant as $m\rightarrow
\infty$, if all of the eigenvalues in Eq.\ (\ref{eigenvalue}) are
distinct \cite{aharonov}.  This will be the case if $(N\eta )/\pi$
is not an integer.

\section{Walk on the line}
The quantum walk on the infinite line can be approached directly or as
the limit of the walk on the cycle as $N$ goes to infinity.  Because
we have just found the eigenstates and eigenvalues for the walk on
the cycle in the previous section, we shall adopt the latter course
here.  In particular, we want to examine what happens when we start
the walk in the state $|0,1\rangle$.  We shall present numerical
results and then follow the approach
developed by Nayak and Vishwanath to study the long time limit of
the probability of being on the edge between the vertices $j$ and
$j+1$.

The probability distribution for the particle after $n$ steps can be
computed in a straightforward manner.  We display the results for the
case $t=r=1/\sqrt{2}$ and the initial state $|0,1\rangle$.  In Figure
1 we have $n=50$ and in Figure 2, $n=1000$.  Note that, as with the
coined quantum walk, these probability distributions are not normal
distributions.  In addition, the region in which the probability
of finding the particle is large is, roughly,in the case of $n=50$, 
between $-35$ and $35$, and in the case $n=1000$, between $-700$
and $700$.  In both cases this corresponds to the high probability
region lying between $-|t|n$ and $|t|n$.  This feature of the
dynamics will be confirmed by our asymptotic analysis.
\begin{figure}
\epsfig{file=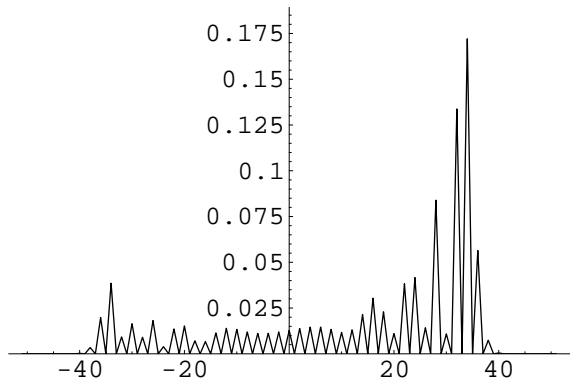}
\caption{Probability distribution for quantum walk after 50
steps}
\end{figure} 
\begin{figure}
\epsfig{file=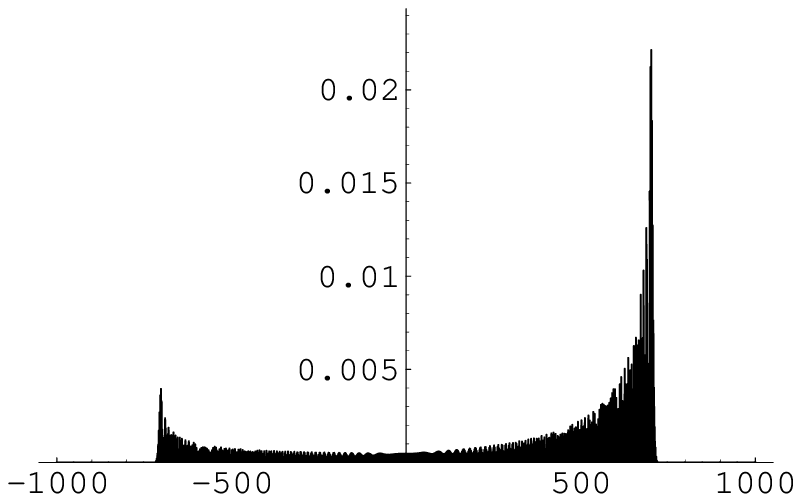}
\caption{Probability distribution for quantum walk after
1000 steps}
\end{figure}

The asymptotic probabilities can be calculated as follows.  
Denoting the eigenstates
corresponding to the eigenvalues $\lambda_{k+}$ and $\lambda_{k-}$ as
$|\psi_{k+}\rangle$ and $|\psi_{k-}\rangle$, respectively, we find that
the wave function of the particle executing the walk, $|\Psi (\tau )
\rangle$ is, after $\tau$ steps,
\begin{eqnarray}
|\Psi (\tau )\rangle & = & \sum_{k=0}^{N-1}\sum_{s=\pm}\lambda_{k,s}^{\tau}
|\psi_{k,s}\rangle\langle\psi_{k,s}|0,1\rangle \nonumber \\
 & = & \frac{1}{\sqrt{N}}\sum_{k=0}^{N-1}(a_{k+}^{(+)\ast}\lambda_{k+}^{\tau}
|\psi_{k+}\rangle +a_{k+}^{(-)\ast}\lambda_{k-}^{tau}|\psi_{k-}\rangle ) ,
\end{eqnarray}
where, as mentioned in the previous paragraph, the initial state is taken 
to be $|0,1\rangle$.  The amplitudes to be in the states $|j.j+1\rangle$
and $|j+1,j\rangle$ at time $\tau$ are given by
\begin{eqnarray}
\label{amplitudes}
\langle j,j+1|\Psi (\tau )\rangle & = & \frac{1}{N}\sum_{k=0}^{N-1}
e^{ij\theta_{k}}(\lambda_{k+}^{\tau}|a_{k+}^{(+)}|^{2}+\lambda_{k-}^{\tau}
|a_{k+}^{(-)}|^{2}) \nonumber \\
\langle j+1,j|\Psi (\tau )\rangle & = & \frac{1}{N}\sum_{k=0}^{N-1}
e^{ij\theta_{k}} (\lambda_{k+}^{\tau} a_{k+}^{(+)\ast}a_{k-}^{(+)}
+\lambda_{k-}^{\tau} a_{k+}^{(-)\ast}a_{k-}^{(-)}) .
\end{eqnarray}
The probability of being on the edge between vertices $j$ and $j+1$ at
time $\tau$, $p(j,j+1;\tau )$, is
\begin{equation}
p(j,j+1;\tau )=|\langle j,j+1|\Psi (\tau )\rangle |^{2}
+|\langle j+1,j|\Psi (\tau )\rangle |^{2} .
\end{equation}

In order to go the the $N\rightarrow\infty$ limit, we need to define
a number of functions of the continuous variable $\theta$ rather than
expressing them as functions of the discrete variable $\theta_{k}$.
We first define
\begin{eqnarray}
C(\theta )= [1-|t|^{2}\cos^{2}(\theta -\eta )]^{1/2} \nonumber \\
S(\theta )=|t|\sin (\theta -\eta ) .
\end{eqnarray}
The eigenvalues also become functions of $\theta$, and we shall express
them as
\begin{equation}
\lambda_{\pm}(\theta )=e^{i\omega_{\pm}(\theta)} ,
\end{equation}
where $0\leq \omega_{+}(\theta )\leq \pi$ and
\begin{equation}
\omega_{+}(\theta )=\tan^{-1}\left(\frac{|t|\cos (\theta -\eta )}
{C(\theta )}\right) ,
\end{equation}
and $-\pi\leq \omega_{-}(\theta ) \leq 0$ and
\begin{equation}
\omega_{-}(\theta )=\tan^{-1}\left(-\frac{|t|\cos (\theta -\eta )}
{C(\theta )}\right) .
\end{equation}
We can now proceed to take the $N\rightarrow\infty$ limit of the sums
appearing in Eqs.\ (\ref{amplitudes}).  For the sums appearing in the
first of these equations we have
\begin{eqnarray}
\label{integrals}
\frac{1}{N}\sum_{k=0}^{N-1}e^{ij\theta_{k}}\lambda_{k+}^{\tau}
|a_{k+}^{(+)}|^{2} &\rightarrow& \int_{0}^{2\pi}d\theta 
e^{i[j\theta +\tau\omega_{+}(\theta )]}\frac{|r|^{2}}{4\pi C(\theta )
[C(\theta )+S(\theta )]} \nonumber \\
\frac{1}{N}\sum_{k=0}^{N-1}e^{ij\theta_{k}}\lambda_{k-}^{\tau}
|a_{k+}^{(-)}|^{2} &\rightarrow& \int_{0}^{2\pi}d\theta 
e^{i[j\theta +\tau\omega_{-}(\theta )]} \nonumber \\
 & & \frac{|r|^{2}}{4\pi C(\theta )[C(\theta )-S(\theta )]} .
\end{eqnarray}
The sums in the second equation become
\begin{eqnarray}
\frac{1}{N}\sum_{k=0}^{N-1} e^{ij\theta_{k}} \lambda_{k+}^{\tau} 
a_{k+}^{(+)\ast}a_{k-}^{(+)} &\rightarrow& -\int_{0}^{2\pi}d\theta
e^{i[j\theta+\tau\omega_{+}(\theta )]}\frac{ir}{4\pi C(\theta )}
\nonumber \\
\frac{1}{N}\sum_{k=0}^{N-1} e^{ij\theta_{k}} \lambda_{k-}^{\tau} 
a_{k+}^{(-)\ast}a_{k-}^{(-)} &\rightarrow& \int_{0}^{2\pi}d\theta
e^{i[j\theta+\tau\omega_{-}(\theta )]}\frac{ir}{4\pi C(\theta )}.
\end{eqnarray}

We are now going to analyze these integrals in the large $\tau$
limit by using the method of stationary phase.  This will be done
in two different ways.  In the first, $j$ will be fixed and $\tau$
will go to infinity.  In the second, we shall set $j=\alpha\tau$,
and then let $\tau$ go to infinity.  Some of the details of this
analysis are given in the appendix.  Here we shall just present the
results.  In the case of fixed $j$ we have that
\begin{eqnarray}
p(j,j+1;\tau )& \sim & \frac{|r|}{\pi\tau |t|} \{ [1+(-1)^{j+\tau}]
\cos^{2}(\tau\mu +\pi /4) \nonumber \\
 & & +[1-(-1)^{j+\tau}]\sin^{2}(\tau\mu +\pi /4)\} ,
\end{eqnarray}
where $0\leq \mu \leq \pi /2$, and
\begin{equation}
\mu =\tan^{-1}\left(\frac{|r|}{|t|}\right) .
\end{equation}
We note that this implies that for any interval located symmetrically
about the origin, the probability of being in that interval goes
like $1/\tau$, whereas for a classical random walk starting at the
origin, it would go like $1/\sqrt{\tau}$.  This implies that, as with
the coined quantum walk, this quantum walk spreads faster than a
classical one.  In the case that $j=\alpha\tau$, we find that there
are stationary phase points only if $\alpha\leq |t|$.  That means 
that for $\alpha > |t|$, $p(j,j+1;\tau)=p(\alpha\tau , \alpha\tau
+1;\tau)$ decreases faster than any inverse power of $\tau$. 
For $\alpha < |t|$ we have that $p(j,j+1;\tau)$ goes like $1/\tau$.  
Therefore, it is most probable that the particle is located in the
region $|j|\leq |t|\tau$, and we can say that the allowed region 
for the particle expands with speed $|t|$.

\section{Relation between quantum walks}
We now have two different quantum walks on the line, the coined walk, where
one moves between vertices, and, what we shall call the edge walk, where
the quantum particle making the walk resides on the edges between the
vertices.  It would be useful to know if the two different walks are related.
In this section we shall show that they are unitarily equivalent.  It should
be emphasized that this result will only be demonstrated for the line, 
whether it holds for more general graphs is not known.  Presently, no
description of a coined walk for a general graph has appeared.

Let us begin by examining the Hilbert spaces for the two different quantum
walks.  The canonical orthonormal basis states of the Hilbert space for the 
coined walk on the line are given by $\{ |j\rangle\otimes |R\rangle , 
|j\rangle\otimes |L\rangle |j\in Z\}$, where the state $|j\rangle$ 
corresponds to the $j$th vertex, and $|r\rangle$ and $|L\rangle$ are the 
coin states.  The Hilbert space in which this walk takes place is just 
$L^{2}(Z)\otimes L^{2}(Z_{2})$.  The canonical orthonormal basis of the Hilbert
space for the edge walk is $\{|j,j+1\rangle ,|j+1,j\rangle |j\in Z \}$,
and the Hilbert space itself is $L^{2}(Z\times Z_{2})$, which is identical
to $L^{2}(Z)\otimes L^{2}(Z_{2})$.

Let us now move to the dynamics.  The unitary operator, $V$, that advances
the coined walk one step is given by
\begin{equation}
V= (S\otimes |R\rangle\langle R|+ S^{\dagger}\otimes |L\rangle\langle L|)
(I\otimes G) ,
\end{equation}
where $G\in U(2)$ is a generalized ``coin-flip'' operator, and is given
by
\begin{eqnarray}
G|R\rangle & = & t|R\rangle + r|L\rangle \nonumber \\
G|L\rangle & = & -r^{\ast}|R\rangle + t^{\ast}|L\rangle .
\end{eqnarray}
The unitary operator, $U$, that advances the edge walk one step was given
in Section 2, and is
\begin{eqnarray}
U|j-1,j\rangle & = & t|j,j+1\rangle +r|j,j-1\rangle \nonumber \\
U|j+1,j\rangle & = & t^{\ast}|j,j-1\rangle -r^{\ast}|j,j+1\rangle .
\end{eqnarray}
Define the unitary operator $\hat{E}$, which takes $L^{2}(Z\times Z_{2})$
into itself, and is given explicitly by
\begin{eqnarray}
\hat{E}|j-1,j\rangle & = & |j\rangle\otimes |R\rangle \nonumber \\
\hat{E}|j+1,j\rangle & = & |j\rangle\otimes |L\rangle .
\end{eqnarray}
We find that 
\begin{equation}
V\hat{E}=\hat{E}U ,
\end{equation}
so that at the level of amplitudes, the two walks are unitarily equivalent.

There is, however, a difference in the probabilities.  In the coined walk,
the probability to be on vertex $j$ is given by combining 
(taking the squares of the magnitudes and adding) the amplitudes
for the states $|j\rangle\otimes |R\rangle$ and $|j\rangle\otimes 
|L\rangle$. Under the mapping $\hat{E}^{-1}$, these states correspond to
states on different edges, $|j-1,j\rangle$ and $|j+1,j\rangle$,
respectively.  However, the probabilities in the edge walk are computed
by combining the amplitudes for being on the same edge, e.g.\ those
for $|j-1,j\rangle$ and $|j,j-1\rangle$.  Therefore, there will be a
difference in the probabilites for the two walks.  This can be seen
explcitly if we examine the probability distribution for the case
$t=r=1/\sqrt{2}$.  We again start in the state $|1,0\rangle$, let
the walk go for $50$ steps, but now compute the probability that
the particle is on a vertex, instead of computing the probability that
it is on an edge.  The result is shown in Figure 3.  By comparing this
figure to Fig. 1, we see that the overall shape of the probability
distributions is similar, but the details are different.
\begin{figure}
\epsfig{file=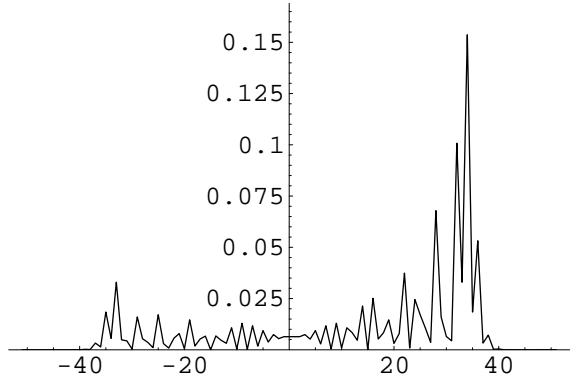}
\caption{Probability distribution for $n=50$, and particle on vertices
instead of edges.}
\end{figure}

\section{Phase shifters}
Going back to the interferometer analogy, we note that we can add a
new element to quantum walks that has no analogue in classical random
walks.  Interferometers are made up of multiports and phase shifters;
a phase shifter imparts a constant phase to a photon that passes through
it.  Suppose we were to put a phase shifter that imparts a phase shift of
$\phi$ just before the $j^{\rm th}$ vertex.  The transition rules for
the states adjacent to this vertex are modified, while the rules for
all other states are unaffected.  In particular, we now have
\begin{eqnarray}
|j-1,j\rangle &\rightarrow & te^{i\phi}|j,j+1\rangle +re^{2i\phi}
|j,j-1\rangle \nonumber \\
|j+1,j\rangle &\rightarrow & -r^{\ast}|j,j+1\rangle +t^{\ast}
e^{i\phi}|j,j-1\rangle .
\end{eqnarray}
Insertion of a phase shifter into an edge can change the properties
of a quantum walk, because it changes how different paths interfere.  

One system that allows us to see their effect on the average 
probability distribution is a modified walk on a cycle.  Suppose that the 
number of vertices is even, and that we put a phase shifter in all of the 
edges whose left end is an even numbered vertex, i.e. every second edge 
has a phase shifter in it.  This system is exactly solvable, and by 
examining its eigenstates, we shall see how the average probability
distributions it gives rise to depend on the value of $\phi$.

The unitary operator that advances this walk one step acts in the following
way if $j$ is even
\begin{eqnarray}
U_{2}|j,j+1\rangle & = & te^{i\phi}|j+1,j+2\rangle +re^{2i\phi}|j+1,j\rangle
\nonumber \\
U_{2}|j+1,j\rangle & = & t^{\ast}|j,j-1\rangle -r^{\ast}|j,j+1\rangle ,
\end{eqnarray}
and if $j$ is odd, then
\begin{eqnarray}
U_{2}|j,j+1\rangle & = & t|j+1,j+2\rangle +r|j+1,j\rangle \nonumber \\
U_{2}|j+1,j\rangle & = & t^{\ast}e^{i\phi}|j,j-1\rangle -r^{\ast}
|j,j+1\rangle .
\end{eqnarray} 
This operator commutes with translations by two steps, i.e.\ with the
operator $T^{2}$.  The eigenstates of $T^{2}$ are given by
\begin{eqnarray}
|u_{k+}^{(e)}\rangle & = & \sqrt{\frac{2}{N}}\sum_{j=0, {\rm even}}^{N-1}
e^{ij\theta_{k}}|j,j+1\rangle \nonumber \\ 
|u_{k+}^{(o)}\rangle & = & \sqrt{\frac{2}{N}}\sum_{j=0, {\rm odd}}^{N-1}
e^{ij\theta_{k}}|j,j+1\rangle \nonumber \\ 
|u_{k-}^{(e)}\rangle & = & \sqrt{\frac{2}{N}}\sum_{j=0, {\rm even}}^{N-1}
e^{ij\theta_{k}}|j+1,j\rangle  \nonumber \\
|u_{k-}^{(o)}\rangle & = & \sqrt{\frac{2}{N}}\sum_{j=0, {\rm odd}}^{N-1}
e^{ij\theta_{k}}|j+1,j\rangle  .
\end{eqnarray}
Each of these states has the eigenvalue $\exp{(-2i\theta_{k})}$.  Eigenstates
of $U_{2}$ are just linear combinations of the above states.  In particular,
expressing the eigenstate of $U_{2}$, $|\psi^{(2)}_{k}\rangle$, as
\begin{equation} 
|\psi^{(2)}_{k}\rangle =a_{k+}|u_{k+}^{(e)}\rangle +a_{k-}|u_{k-}^{(e)}
\rangle + b_{k+}|u_{k+}^{(o)}\rangle +b_{k-}|u_{k-}^{(o)}\rangle ,
\end{equation}
the eigenvalue equation $U_{2}|\psi^{(2)}_{k}\rangle =\lambda
|\psi^{(2)}_{k}\rangle$ becomes
\begin{equation}
\left( \begin{array}{cccc} 0 & 0 & -r^{\ast} & te^{-i\theta_{k}} \\
0 & 0 & t^{\ast}e^{i\theta_{k}} & r \\ re^{2i\phi} & t^{\ast}
e^{i(\theta_{k}+\phi)} & 0 & 0 \\ te^{i(\phi -\theta_{k})} & 
-r^{\ast} & 0 & 0 \end{array}\right)
\left( \begin{array}{c} a_{k+} \\ b_{k-} \\ a_{k-} \\ b_{k+} \end{array}
\right) = \lambda \left( \begin{array}{c} a_{k+} \\ b_{k-} \\ a_{k-} \\ 
b_{k+} \end{array} \right) .
\end{equation}
The eigenvalues satisfy the equation
\begin{equation}
\lambda^{4}+\lambda^{2}[|r|^{2}(1+e^{2i\phi})-e^{i\phi}(t^{\ast 2}
e^{2i\theta_{k}}+t^{2}e^{-2i\theta_{k}})]+e^{2i\phi} =0 .
\end{equation}

The eigenstates of this system no longer give rise to constant 
probability distributions; the probabilities of being on an even
edge (an edge whose left-most vertex is even) and an odd edge (an
edge whose left-most vertex is odd) are, in general, different.
If $\phi =0$, these probabilities are the same, but if $\phi =\pi /2$,
then this is no longer the case.  In the latter case we find that
\begin{equation}
\lambda^{2}=i|t|^{2}\cos(2\theta_{k}-2\eta )\pm [1-|t|^{4}\cos^{2}
(2\theta_{k}-2\eta )]^{1/2} .
\end{equation}
Choosing the plus sign in the above equation, we find that for an 
eigenfunction corresponding to $\theta_{k}-\eta =\pi /4$, we have 
for the ratio of the probability of being on an even edge to being 
on an odd one
\begin{equation}
\frac{p_{even}}{p_{odd}}=\frac{1+|r|^{2}}{1-|r|^{2}} .
\end{equation}
If $\theta_{k}-\eta$ cannot be exactly equal to $\pi /4$ because
of the values of $N$ or $\eta$, then for $\theta_{k}-\eta$ close
to $\pi /4$ the ratio of even-edge to odd-edge probabilities will
be approximately given by the above equation.  
This ratio is not generally equal to one, which means that the
average probability distribution to which a given initial state
converges will not be constant.  The introduction of the phase
shifters has changed the character of the quantum walk.

These changes can also be seen by calculating the probability
distributions after a finite number of steps.  This is done for
$50$ steps and for the case $t=r=1/\sqrt{2}$ and 
$\phi =\pi /2,\: \pi /3$ in the following figures.  The initial
state is, as before, $|0,1\rangle$.  These can be
compared to Fig.\ 1, which corresponds to the case $\phi =0$.
It can be seen that the introduction of the phase shifter greatly
changes the character of the probability distribution.  Note that
particularly for the case of $\phi =\pi /2$, the size of the 
region in which it is very likely that the particle will be found
is smaller than when $\phi =0$.  For a small number of steps, it is
easy to verify by hand that destructive interference in the 
$\phi =\pi /2$ case makes the walk spread more slowly than when
$\phi =0$, and the numerical results indicate that this feature
persists for at least $50$ steps.

If we extend this walk to the infinite line, the difference caused by
the phase shifters can be seen in the asymptotic behavior.  For both
$\phi =0$ and $\phi = \pi /2$ the size of the region in which it is
most likely to find the particle grows linearly with the number of 
steps, but the ``speed'' is different.  We saw that in the case $\phi =0$
the probability distribution $p(j,j+1:\tau )$ falls off rapidly for
$|j|>|t|\tau$.  If $\phi = \pi /2$, it falls off rapidly for $|j|>
|t|^{2}\tau$, which means that the size of the high probability region
is smaller in this case.  We see yet again, that phase effects, which do
not exist in classical random walks, can significantly influence the
behavior of quantum walks.
\begin{figure}
\epsfig{file=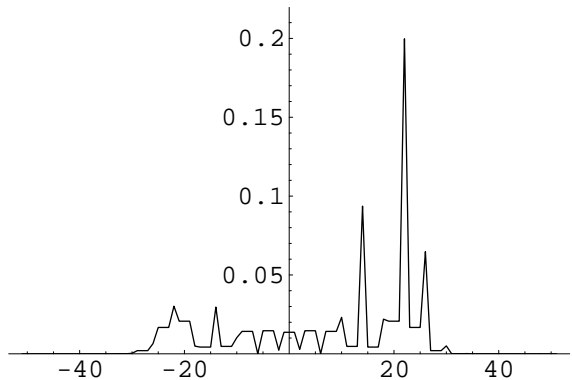}
\caption{Probability distribution for $n=50$ and $\phi =\pi /2$.}
\end{figure}
\begin{figure}
\epsfig{file=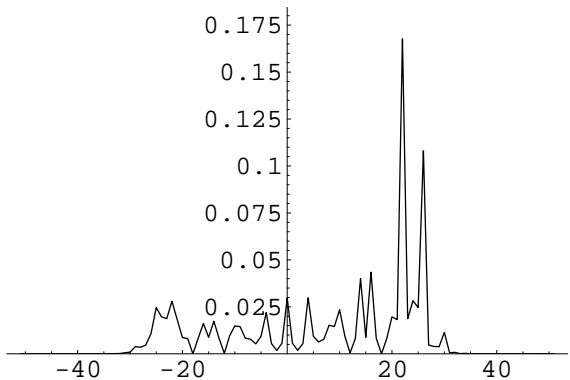}
\caption{Probability distribution for $n=50$ and $\phi =\pi /3$.}
\end{figure}

\section{Probability current}
In standard quantum mechanics, it is possible to define a probability
current density.  In one dimension, if the wave function of the particle 
is $\psi (x,t)$, then the probability current density is given by
\begin{equation}
j(x,t)=\frac{1}{2im}\left[ \psi^{\ast}(x,t)\frac{d}{dx}\psi (x,t)-\psi (x,t)
\frac{d}{dx}\psi^{\ast}(x,t)\right] ,
\end{equation}
where $m$ is the mass of the particle, and we are using units in
which $\hbar =1$.  This current has the property that
\begin{equation}
\label{divj}
\frac{\partial}{\partial t}\int_{x_{1}}^{x_{2}}|\psi (x,t)|^{2}=
-[j(x_{2},t)-j(x_{1},t)] ,
\end{equation}
that is, the change in the probability of the particle being in a
particular region is given by the net flow of probability into the
region.  For an eigenstate of the Hamiltonian, the probability
density $|\psi (x,t)|^{2}$ is independent of time, so that the
probability current density is a constant.  We would like to show 
that there is a quantity similar to the probability current density 
for quantum walks.

Suppose that the state of the walk on the cycle is given by
\begin{equation}
|\Psi\rangle =\sum_{j=0}^{N-1}(c_{j,j+1}|j,j+1\rangle +c_{j+1,j}
|j+1,j\rangle ) .
\end{equation}
Define the probability current at the $k^{\rm th}$ vertex to be
\begin{equation}
J_{k}=(c^{\ast}_{k+1,k}\: c^{\ast}_{k-1,k})\left(\begin{array}{cc}
|t|^{2} & tr \\ t^{\ast}r^{\ast} & -|t|^{2} \end{array}\right)
\left(\begin{array}{c} c_{k+1,k} \\ c_{k-1,k} \end{array}\right) .
\end{equation}
We find that, if $\Delta P_{k,k+1}$ is the change in the probability
of being on the edge between vertices $k$ and $k+1$ in one step of
the walk, then
\begin{equation}
\label{deltaj}
\Delta P_{k,k+1}=J_{k+1}-J_{k} ,
\end{equation}
which is the discrete analogue of Eq.\ (\ref{divj}).  For an eigenstate
of the walk, this current, $J_{k}$, must be independent of $k$.  The
above equation holds even if the transmission and reflection amplitudes
are different for each beam splitter, i.e.\ they depend on $k$, and
also if phase shifters are present.  In that case, if $t_{k}$ is the
transmission amplitude at vertex $k$, $r_{k}$ the reflection
amplitude, and $\phi_{k}$ the phase shift of the phase shifter just
to the left of vertex $k$, the current at this vertex is give by
\begin{equation}
J_{k}=(c^{\ast}_{k+1,k}\: c^{\ast}_{k-1,k})\left(\begin{array}{cc}
|t_{k}|^{2} & tre^{i\phi_{k}} \\ t^{\ast}r^{\ast}e^{-i\phi_{k}} & 
-|t_{k}|^{2} \end{array}\right)
\left(\begin{array}{c} c_{k+1,k} \\ c_{k-1,k} \end{array}\right) .
\end{equation}

 We can use this fact to demonstrate a general property of eigenstates of
certain kinds of walks on a line.  Suppose that all of the beam splitters
located at vertices $k<0$ and $k>N$ have transmission amplitude $t=1$ and 
reflection amplitude $r=0$.  The beam splitters for $0\leq k \leq N$ can 
have any value of transmission amplitude and reflection amplitude, and 
these values can vary from vertex to vertex.  We shall refer to the
vertices between $0$ and $N$ as the scattering region.  The problem 
we are considering is analogous to the scattering of a particle moving
in one dimension off of a potential, which is nonzero only in a bounded
interval.  From the point of view of a walk, we may be interested in a
walk that starts to the left of the scattering region in a right-moving
state, and finding out how long it takes to get through the scattering
region.

The eigenstates of this type of walk are of two types.  The first of a
particle coming in from the left, its reflected amplitude, and a 
transmitted amplitude to the right of the scattering region.  The
second consists of a particle coming in from the right, its reflected
amplitude, and a transmitted amplitude to the left of the scattering
region.  We shall consider the first type, which can be expressed as
\begin{equation}
|\Psi\rangle = \sum_{j=-\infty}^{N-1}(c_{j,j+1}|j,j+1\rangle +c_{j+1,j}
|j+1,j\rangle )+\sum_{j=N}^{\infty}c_{j,j+1}|j,j+1\rangle .
\end{equation}
Setting the eigenvalue, $\lambda$, equal to $\exp (-i\theta )$, the 
equation $U|\Psi\rangle =\lambda |\Psi\rangle$ gives us that
\begin{eqnarray}
c_{j,j+1}=e^{i(j+1)\theta}c_{-1,0} & \hspace {1cm}& {\rm for}  j\leq -1
\nonumber \\
c_{j+1,j}=e^{-i(j+1)\theta}c_{0,-1} & \hspace{1cm}& {\rm for} j\leq -1
\nonumber \\
c_{j,j+1}=e^{i(j-N)\theta}c_{N,N+1} & \hspace{1cm} & {\rm for} 
j\geq N .
\end{eqnarray}
The amplitude $c_{-1,0}$ can be thought of as the amplitude of the incoming
wave, $c_{0,-1}$ the amplitude of the reflected wave, and $c_{N,N+1}$
the amplitude of the transmitted wave.  We can find a condition that
these quantities must satisfy, if we  make use of the fact that 
$J_{-1}=J_{N+1}$.  This gives us that
\begin{equation}
|c_{0,-1}|^{2}+|c_{N,N+1}|^{2}=|c_{-1,0}|^{2} ,
\end{equation}
where we have used the fact that $|c_{-2,-1}|^{2}=|c_{-1,0}|^{2}$.
Defining the reflection coefficient of the scattering region to be
$R=|c_{0,-1}|^{2}/|c_{-1,0}|^{2}$ and the transmission coefficient to
be $T=|c_{N,N+1}|^{2}/|c_{-1,0}|^{2}$, then the above equation can
be expressed as $R+T=1$.

\section{Vertices with more than two edges}
So far we have only considered vertices at which two edges meet, but
if we are to construct graphs more complicated than lines, we need
to see how a vertex with more that two edges emanating from it behaves.
We shall look at two examples, one with three edges, and another with
an arbitrary number.

A vertex with three edges emanating from it, which is inspired by the
optical multiport known as the tritter \cite{zeilinger}, can be 
described as follows.  Let us label the vertex with three edges meeting
at it by $O$, and the opposite ends of the edges by $A$, $B$, and $C$.
The ingoing states for this vertex are $|AO\rangle$, $|BO\rangle$,
and $|CO\rangle$, and the outgoing states are $|OA\rangle$, $|OB\rangle$,
and $|OC\rangle$.  Setting $z=\exp{(2\pi i/3)}$, we have for the transition
rules
\begin{eqnarray}
|AO\rangle &\rightarrow &\frac{1}{\sqrt{3}}(|OA\rangle +|OB\rangle
+|OC\rangle ) \nonumber \\
|BO\rangle &\rightarrow & \frac{1}{\sqrt{3}}(z^{\ast}|OA\rangle 
+|OB\rangle +z|OC\rangle ) \nonumber \\
|CO\rangle &\rightarrow &\frac{1}{\sqrt{3}}(z^{\ast}|OA\rangle 
+z|OB\rangle +|OC\rangle ) .
\end{eqnarray}
This vertex has the property that an incoming particle is equally 
likely to exit through each edge.  However, note that because the 
incoming states from different edges behave differently with regard
to their phases, the use of this vertex requires the labelling of
edges.  In this particular case, only one of the edges needs to
be labelled.  If we attach a label to $AO$, we interpret it to 
mean that if the input state is along either of the other two 
edges, then the output with a phase factor of $z^{\ast}$ is along
the labelled edge.  For any edge, if the input state is along this
edge, the part of the output state along the same edge has a
phase factor of one.

It is possible to define vertices that do not require the labelling
of edges, though the cost is that the probabilities of exiting through
each of the edges are no longer the same.  This particular kind of
vertex is very closely related to the quantum coin used in the
walk on the hypercube \cite{moore,kempe}.  Let the vertex at which
all of the edges meet be labelled by $O$, and the opposite ends of
the edges be labelled by the numbers $1$ through $n$.  For any 
input state, $|kO\rangle$, where $k$ is an integer between $1$ and
$n$, the transition rule is that the amplitude to go the output
state $|Ok\rangle$ is $r$, and the amplitude to go to any other
output state is $t$.  That is, the amplitude to be reflected is $r$,
and the amplitude to be transmitted through any of the other edges
is $t$.  Unitarity places two conditions on these amplitudes
\begin{eqnarray}
(n-1)|t|^{2}+|r|^{2}=1 \nonumber \\
(n-2)|t|^{2}+r^{\ast}t+t^{\ast}r =0 .
\end{eqnarray}
As an example, for the case $n=3$, possible values of $r$ and $t$
are $r=-1/3$ and $t=2/3$.  Because each of the edges in this
vertex behaves in the same way, they are equivalent to each other
and no labelling is necessary.
 
In order to construct a walk for a general graph, one chooses a unitary
operator for each vertex, i.e.\ one that maps the states coming into
a vertex to states leaving the same vertex.  One step of the walk consists
of the combined effect of all of these operations; the overall unitary
operator,$U$, that advances the walk one step is constructed from the local
operators for each vertex.  Explicitly, the edge state $|ab\rangle$,
which can be interpreted as going from vertex $a$ to vertex $b$, will
go to the state $U_{b}|ab\rangle$ after one step, where $U_{b}$ is
the operator corresponding to vertex $b$.  This prescription guarantees
that the overall operation is unitary, in particular, $U$ acting on
any other edge state $|cd\rangle$ will give a state orthogonal to
$U|ab\rangle$.  If $d=b$, then $|ab\rangle$ and $|cd\rangle$ will be
mapped onto the same set of states (the states leaving vertex $b$),
but the unitarity of $U_{b}$ will ensure that $U|ab\rangle$ and
$U|cd\rangle$ are orthogonal.  If $d\neq b$, then $U$ maps $|ab\rangle$
and $|cd\rangle$ onto different sets of states, and the so results are
then orthogonal.  Therefore, as the edge states make up an orthonormal
basis of the Hilbert space in which the walk occurs, and $U$ maps 
this basis to another orthonormal basis, it is unitary.

\section{Conclusion}
Quantum walks on graphs seem promising for the development of algorithms,
because they spread over a graph faster than does a classical walk,
and can thereby explore the structure of the graph faster than can
a classical random walk.  Here we have discussed a discrete quantum walk 
that is based on an analogy to optical interferomenters.  The vertices
act as optical multiports and phase shifters can be inserted into
the edges.  As we have seen, the behavior of this type of walk depends
on both types of elements.  The advantage of this type of quantum walk
is that it can easily be defined for any graph.  

In this paper, we have, for the most part, confined our attention to the
quantum walk on the line.  We found that the probability distribution of
the particle making the walk spreads linearly with the number of steps, 
and with a ``speed'' given by $|t|$.  In addition, there is a probability
current, whose ``divergence'' gives the probability flowing into an edge.
This allowed us to define reflection and transmission coefficients for
one-dimensional graphs.

The extension of these results to more general graphs is clearly the
next step.  It has been shown how definine a quantum walk for any 
graph, but the properties of these walks have yet to be explored. 

\section*{Acknowledgments}
This research was supported in part by the National Science Foundation
under grant number PHY-0139692 and by the Office of Naval Research under
grant number N00014-94J-1233.  We would also like to thank Jozef
\v{S}kvar\v{c}ek for preliminary numerical work on this problem. 

\section*{Appendix}
Here we want to explicitly show how the asymptotic analysis on the integrals
in section III.  We shall show how to do the analysis for the first 
integral in Eq.\ (\ref{integrals}), and the procedure for the others is
similar.

We shall first consider the case when $j$ is fixed.  Then the stationary
phase points are the solutions of the equation $\omega^{\prime}_{+}
(\theta )=0$, and we find the two solutions $\theta = \eta$ and $\theta
=\eta +\pi$.  We also find that $\omega_{+}^{\prime\prime}(\eta )=
|t|/|r|$ and $\omega_{+}^{\prime\prime}(\eta +\pi )=-|t|/|r|$.  In
addition, $\omega_{+}(\eta )=\mu$ and $\omega_{+}(\eta +\pi )=\pi -\mu$.
Inserting these values into the standard formula for stationary
phase \cite{bender}, we find that
\begin{eqnarray}
\int_{0}^{2\pi}d\theta e^{i[j\theta +\tau\omega_{+}(\theta )]}\frac{|r|^{2}}
{4\pi C(\theta )[C(\theta )+S(\theta )]} &\sim & \frac{1}{2}\left( \frac
{|r|}{2\pi \tau |t|}\right)^{1/2}e^{ij\eta} \nonumber \\
(e^{i(\tau\mu+\pi /4)}+(-1)^{j+\tau}e^{-i(\tau\mu+\pi /4)}) .
\end{eqnarray}

Let us now consider the case when $j=\alpha\tau$.  The stationary
phase points are now given by the solutions of
\begin{equation}
\label{stationary}
\omega_{+}^{\prime}(\theta )=\frac{|t|\sin (\theta -\eta )}{[1-|t|^{2}
\cos^{2}(\theta -\eta )]^{1/2}}=-\alpha .
\end{equation}
For this equation to have any solutions, we find that $\alpha \leq |t|$.
If this condition is not satisfied, there are no stationary phase points,
and the integral decreases faster than any inverse power of $\tau$.
If $\alpha <|t|$, then there are two solutions, and they satisfy the 
conditions $\pi \leq \theta -\eta \leq 2\pi$ and
\begin{equation}
\sin^{2}(\theta -\eta )=\frac{(\alpha |t|)^{2}}{(1-\alpha^{2})|t|^{2}} .
\end{equation}
Explicitly, if $\gamma$ lies between $0$ and $\pi /2$ and satisfies
\begin{equation}
\sin^{2}(\gamma )=\frac{(\alpha |t|)^{2}}{(1-\alpha^{2})|t|^{2}} ,
\end{equation}
then the two solutions to Eq.\ (\ref{stationary}) are $\theta_{1}=\eta
+\gamma +\pi$ and $\theta_{2}=\eta +2\pi -\gamma$.  We find that
\begin{eqnarray}
\omega_{+}^{\prime\prime}(\theta_{1}) & = & -\frac{1}{|r|}(|t|^{2}-
\alpha^{2})^{1/2}(1-\alpha^{2}) \nonumber \\
\omega_{+}^{\prime\prime}(\theta_{2}) & = & \frac{1}{|r|}(|t|^{2}-
\alpha^{2})^{1/2}(1-\alpha^{2}) ,
\end{eqnarray}
and $\omega_{+}(\theta_{1})=\pi -\nu$ and $\omega_{+}(\theta_{2})=\nu$,
where $0\leq \nu \leq \pi /2$ and
\begin{equation}
\nu = \tan^{-1}\left(\frac{|r|}{(|t|^{2}-\alpha^{2})^{1/2}}\right) .
\end{equation}
Finally for the integral we find that
\begin{eqnarray}
\int_{0}^{2\pi}d\theta e^{i\tau [\alpha\theta +\omega_{+}(\theta )]}
\frac{|r|^{2}}{4\pi C(\theta )[C(\theta )+S(\theta )]} \nonumber \\
\sim \frac{1}{2(1-\alpha )}\left[\frac{|r|(1-\alpha^{2})}{2\pi\tau
(|t|^{2}-\alpha^{2})^{1/2}}\right]^{1/2}e^{i\alpha\tau\eta}[(-1)^{\tau}
e^{i\tau [\alpha (\pi +\gamma )-\nu]-i\pi /4} \nonumber \\
+e^{i\tau [\alpha (2\pi -\gamma )+\nu]+i\pi /4}] .
\end{eqnarray}
Finding the asymptotic form of the other integrals in a similar fashion,
we have for $j=\alpha\tau$ and $\alpha <|t|$
\begin{eqnarray}
p(j,j+1;\tau ) & \sim & \frac{|r|}{\pi\tau (|t|^{2}-\alpha^{2})^{1/2}
(1-\alpha)} [1+\alpha (-1)^{\tau}\cos (\pi\alpha\tau )]
\nonumber \\
 & &\{ 1+(-1)^{\tau}\sin [2\tau (\alpha\gamma -\nu )-\pi\alpha\tau ] \} .
\end{eqnarray}

\bibliographystyle{unsrt}

\end{document}